\documentclass[runningheads]{llncs}
\usepackage[T1]{fontenc}
\usepackage{graphicx}
\usepackage{booktabs}
\usepackage{multirow}
\usepackage{amssymb}
\usepackage{xcolor}
\usepackage[colorlinks=true, linkcolor=blue, urlcolor=blue, citecolor=blue, anchorcolor=blue]{hyperref}

\begin{document}

\title{Conditional Temporal Attention Networks for Neonatal Cortical Surface Reconstruction}
\titlerunning{Conditional Temporal Attention Networks}
%
\author{Qiang Ma\inst{1}\and
Liu Li\inst{1}\and
Vanessa Kyriakopoulou\inst{2}\and
Joseph Hajnal\inst{2}\and\\
Emma C. Robinson\inst{2}\and
Bernhard Kainz\inst{1,2,3}\and
Daniel Rueckert\inst{1,4}
}


\authorrunning{Q. Ma et al.}

\institute{BioMedIA, Department of Computing, Imperial College London, UK \\
\email{q.ma20@imperial.ac.uk}
\and
King's College London, UK
\and
FAU Erlangen–N\"urnberg, Germany
\and
Klinikum rechts der Isar, Technical University of Munich, Germany
}

\maketitle              
\begin{abstract}
Cortical surface reconstruction plays a fundamental role in modeling the rapid brain development during the perinatal period. In this work, we propose Conditional Temporal Attention Network (CoTAN), a fast end-to-end framework for diffeomorphic neonatal cortical surface reconstruction. CoTAN predicts multi-resolution stationary velocity fields (SVF) from neonatal brain magnetic resonance images (MRI). Instead of integrating multiple SVFs, CoTAN introduces attention mechanisms to learn a conditional time-varying velocity field (CTVF) by computing the weighted sum of all SVFs at each integration step. The importance of each SVF, which is estimated by learned attention maps, is conditioned on the age of the neonates and varies with the time step of integration. The proposed CTVF defines a diffeomorphic surface deformation, which reduces mesh self-intersection errors effectively. It only requires 0.21 seconds to deform an initial template mesh to cortical white matter and pial surfaces for each brain hemisphere. CoTAN is validated on the Developing Human Connectome Project (dHCP) dataset with 877 3D brain MR images acquired from preterm and term born neonates. Compared to state-of-the-art baselines, CoTAN achieves superior performance with only 0.12$\pm$0.03mm geometric error and 0.07$\pm$0.03\% self-intersecting faces. The visualization of our attention maps illustrates that CoTAN indeed learns coarse-to-fine surface deformations automatically without intermediate supervision.

\end{abstract}
\section{Introduction}

Cortical surface reconstruction aims to extract 3D meshes of inner (white matter) and outer (pial) surfaces of the cerebral cortex from brain magnetic resonance images (MRI). These surfaces provide both 3D visualization and estimation of morphological features for the cortex~\cite{dale1999cortical1,fischl2000thickness,fischl1999cortical2}. 
In addition to accurately representing the highly folded cortex, the cortical surfaces of each hemisphere are required to be closed manifolds and topologically homeomorphic to a sphere~\cite{dale1999cortical1}. Traditional neuroimage analysis pipelines~\cite{fischl2012freesurfer,glasser2013hcp,makropoulos2018dhcp,shattuck2002brainsuite} such as FreeSurfer~\cite{fischl2012freesurfer} comprise a series of processing steps to extract cortical surfaces from brain MRI. These pipelines provide an invaluable service to the research community. However, the current implementations have limited accuracy and require several hours to process a single MRI scan.

With the recent advances of geometric deep leaning \cite{wang2018pixel2mesh,wickramasinghe2020voxel2mesh}, a growing number of fast learning-based approaches have been proposed to learn either implicit surface representation~\cite{cruz2021deepcsr,gopinath2021segrecon} or explicit mesh deformation~\cite{bongratz2022vox2cortex,hoopes2021topofit,lebrat2021corticalflow,ma2022cortexode,ma2021pialnn,santa2022cfpp} for cortical surface reconstruction. These approaches enhance the accuracy and reduce the processing time to a few seconds for a single subject.
Recent studies focus on generating manifold cortical meshes~\cite{lebrat2021corticalflow,ma2022cortexode,santa2022cfpp} and preventing mesh self-intersections by learning diffeomorphic deformations, which have been widely adopted in medical image registration~\cite{ashburner2007dartel,balakrishnan2019voxelmorph,beg2005lddmm}. The basic idea is to learn stationary velocity fields (SVF) to deform an initial mesh template to target surfaces. Since a single SVF has limited representation capacity, several  approaches~\cite{gupta2020nmf,lebrat2021corticalflow,santa2022cfpp,sun2022topology} have proposed to train multiple neural networks to predict a sequence of SVFs for coarse-to-fine surface deformation. This improves the geometric accuracy but increases the computational burden for both training and inference.

\begin{figure}[t]
\centering
\begin{minipage}[c]{0.52\textwidth}
\centering
\includegraphics[width=1.0\textwidth]{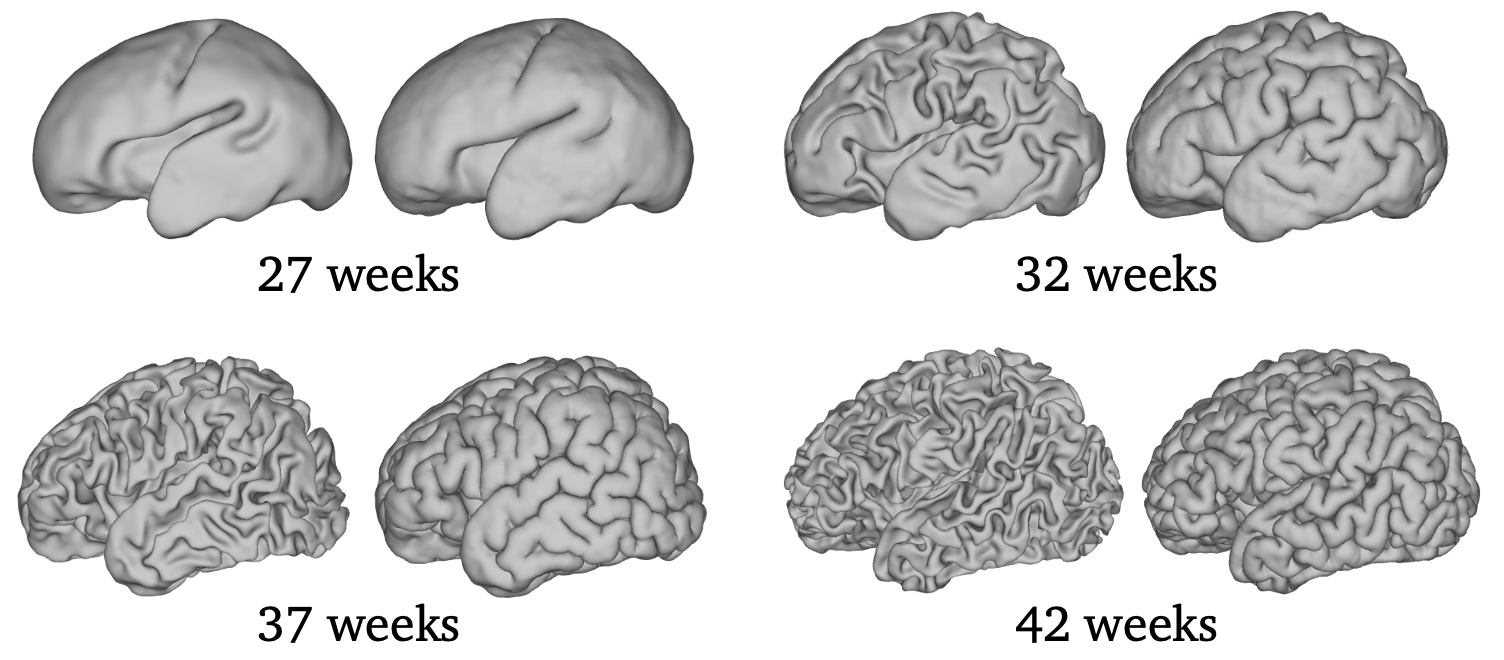}
\caption{Neonatal cortical surfaces at different post-menstrual ages.}
\label{fig:age}
\end{minipage}
\hfill
\begin{minipage}[c]{0.45\textwidth}
\centering
\includegraphics[width=1.0\textwidth]{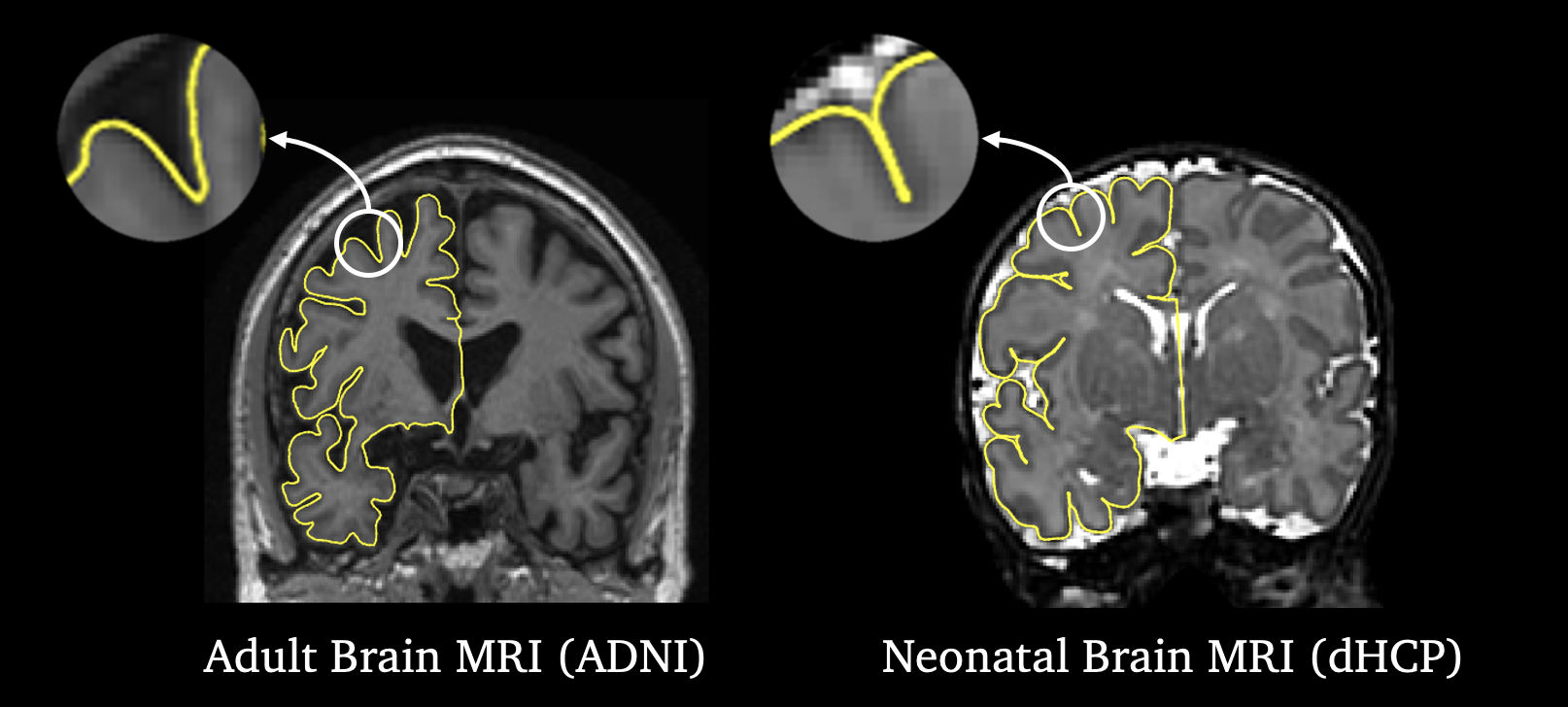}
\caption{Left: adult brain MRI from the ADNI dataset~\cite{jack2008adni}. Right: neonatal brain MRI from the dHCP dataset~\cite{edwards2022dhcp}.}
\label{fig:sulci}
\end{minipage}
\end{figure}

Cortical surface reconstruction plays an essential role in modeling and quantifying the brain development in fetal and neonatal neuroimaging studies such as the Developing Human Connectome Project (dHCP)~\cite{edwards2022dhcp,makropoulos2018dhcp}. However, most learning-based approaches so far rely on adult MRI as training data~\cite{glasser2013hcp,jack2008adni,marcus2007oasis}.
Compared to adult data, neonatal brain MR images have lower resolution and contrast due to a smaller region of interest and the use of, \emph{e.g.}, fast imaging sequences with sparse acquisition to minimize head motion artifacts of unsedated infants~\cite{cordero2018motion,hughes2017dedicated}. 
Besides, the rapid growth and continuously increasing complexity of the cortex during the perinatal period lead to considerable variation in shape and scale between neonatal cortical surfaces at different post-menstrual ages (PMA) (see Figure \ref{fig:age}). Moreover, since the neonatal head is still small, the cortical sulci of term-born neonates are much narrower than adults as shown in Figure \ref{fig:sulci}. Hence, the neonatal pial surfaces are prone to be affected by partial volume effects and more likely to produce surface self-intersections.

\textbf{Contribution.} In this work, we present Conditional Temporal Attention Network (CoTAN). CoTAN adopts attention mechanism~\cite{hu2018squeeze,roy2018concurrent} to learn a conditional time-varying velocity field (CTVF) for neonatal cortical surface reconstruction. Given an input brain MR image, CoTAN first predicts multiple SVFs at different resolutions. Rather than integrating all SVFs as~\cite{gupta2020nmf,lebrat2021corticalflow,santa2022cfpp}, CoTAN learns conditional temporal attention maps to attend to specific SVFs for different time steps of integration and PMA of neonates. The CTVF is represented by the weighted sum of learned SVFs, and thus a single CoTAN model is sufficient to model the large deformation and variation of neonatal cortical surfaces. The evaluation on the dHCP neonatal dataset~\cite{edwards2022dhcp} verifies that CoTAN performs better in geometric accuracy, mesh quality and computational efficiency than state-of-the-art methods. The visualization of attention maps indicates that CoTAN learns coarse-to-fine deformations automatically without intermediate constraints. Our source code is released publicly at \url{https://github.com/m-qiang/CoTAN}.

\section{Method}
\noindent\textbf{Diffeomorphic Surface Deformation. }
We define the diffeomorphic surface deformation $\phi_t:\mathbb{R}^3\times\mathbb{R}\rightarrow\mathbb{R}^3$ as a flow ordinary differential equation (ODE) following previous work~\cite{ashburner2007dartel,balakrishnan2019voxelmorph,beg2005lddmm,gupta2020nmf,lebrat2021corticalflow,ma2022cortexode,santa2022cfpp}:
\begin{equation}\label{eq:flow}
\frac{\partial}{\partial t}\phi_t = v_t(\phi_t),~\phi_0=Id,~t\in[0,T],
\end{equation}
where $v_t$ is a time-varying velocity field (TVF) and $Id$ is the identity mapping. Given an initial surface $S_0\subset\mathbb{R}^3$ with points $x_0\in S_0$, we define $x_t:=\phi_t(x_0)$ as the trajectories of the points on the deformable surface $S_t=\phi_t(S_0)$ for $t\in[0,T]$. Then the flow equation~(\ref{eq:flow}) can be rewritten as $\frac{d}{d t}{x}_t = v_t(x_t)$ with initial value $x_0$. By the existence and uniqueness theorem for ODE solutions~\cite{gupta2020nmf}, if $v_t(x)$ is Lipschitz continuous with respect to $x$, the trajectories $x_t$ will not intersect with each other, so that the surface self-intersections can be prevented effectively. By integrating the ODE~(\ref{eq:flow}), we obtain a diffeomorphism $\phi_T$ that deforms $S_0$ to a manifold surface $S_T$, on which the points are $x_T=\phi_T(x_0)=x_0+\int_0^Tv_t(x_t)dt$.
\newline

\begin{figure}[t]
\centering
\includegraphics[width=1.0\textwidth]{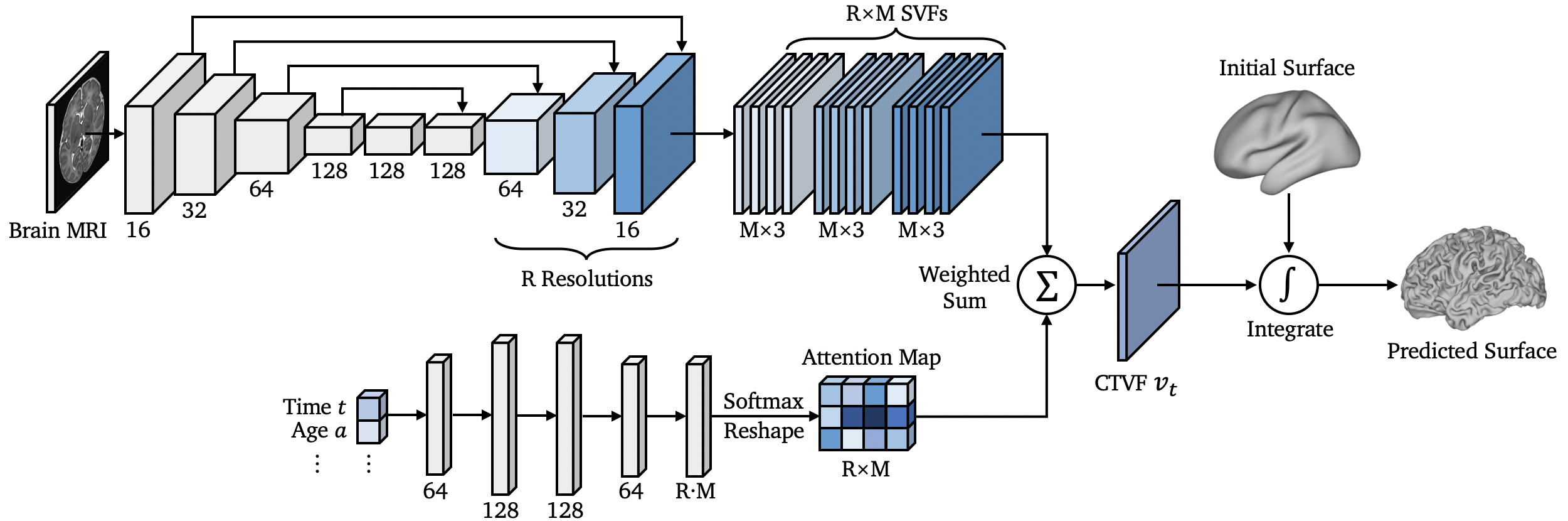}
\caption{The architecture of the proposed CoTAN framework. Given an input 3D brain MRI, CoTAN uses a U-Net to predict $M$ SVFs for each resolution level $R$. An attention map is learned to focus on specific SVFs varying with the input integration time $t$ and conditioned on the age $a$ of the neonatal subjects. Further conditioning could be achieved with minimal effort, \emph{e.g.}, biological sex, diagnoses, etc. For each time step $t$, the CTVF $v_t$ is represented by the weighted sum of all $R\times M$ SVFs. By integrating the CTVF, CoTAN deforms an input initial surface to the predicted cortical surface.}
\label{fig:cotan}
\end{figure}

\noindent\textbf{Conditional Temporal Attention Network (CoTAN). }
An overview of the CoTAN architecture is shown in Figure \ref{fig:cotan}. 
CoTAN first predicts multiple SVFs given a 3D brain MRI volume. A 3D U-Net~\cite{ronneberger2015unet} is used to extract feature maps with $R$ resolution levels, each of which scales the input size by the factor of $2^{r-R}$ for $r=1,...,R$. Then we upsample the multi-scale feature maps and learn $M$ volumetric SVFs for each resolution. 
Let $\mathbf{V}$ denote all $R\times M$ discrete SVFs. The continuous multi-resolution SVFs $\mathbf{v}:\mathbb{R}^3\rightarrow\mathbb{R}^{R\times M\times 3}$ can be obtained by $\mathbf{v}(x)=\mathrm{Lerp}(x, \mathbf{V})$, where $\mathrm{Lerp}(\cdot)$ is the trilinear interpolation function. Each element $\mathbf{v}^{r,m}:\mathbb{R}^3\rightarrow\mathbb{R}^3$ is an SVF for $r=1,...,R$ and $m=1,...,M$. Note that $\mathbf{v}(x)$ is Lipschitz continuous since $\mathrm{Lerp}(\cdot)$ is continuous and piece-wise linear.

CoTAN adopts a channel-wise attention mechanism~\cite{hu2018squeeze,roy2018concurrent} to focus on specific SVFs since it is time-consuming to integrate all $R\times M$ SVFs~\cite{gupta2020nmf,lebrat2021corticalflow,santa2022cfpp}. The attention is conditioned on both integration time $t\in[0,T]$ and information about the subject. To model the high variation between infant brains, we consider the post-menstrual ages (PMA) $a\in\mathbb{R}$ of the neonates at scan time as the conditioning variable in this work. Note that we do not use a self-attention module~\cite{dosovitskiy2020vit,vaswani2017attention} to learn key and query pairs. Instead, we learn a probability attention map to measure the importance of each SVF.
More precisely, as shown in Figure \ref{fig:cotan}, we use a fully connected network (FCN) to encode the input time $t$ and PMA $a$ into a $(R\cdot M) \times 1$ feature vector. After reshaping and softmax activation, the FCN learns conditional temporal attention maps $\mathbf{p}(t,a)\in\mathbb{R}^{R\times M}$ which satisfy $\sum_{r=1}^R\sum_{m=1}^M\mathbf{p}^{r,m}(t,a)=1$ for any $t$ and $a$. Then a \textit{conditional time-varying velocity field} (CTVF) is predicted by computing the weighted sum of all SVFs:
\begin{equation}\label{eq:ctvf}
v_t(x;a)=\sum\nolimits_{r=1}^R\sum\nolimits_{m=1}^M \mathbf{p}^{r,m}(t,a)\cdot \mathbf{v}^{r,m}(x).
\end{equation}
The CTVF is adaptive to the integration time and the age of subjects, which can handle the large deformation and variation for neonatal cortical surfaces. Such an attention mechanism encourages CoTAN to learn coarse-to-fine surface deformation by attending to SVFs at different resolutions.

To deform the initial surface $S_0$ to the target surface, we integrate the flow ODE (\ref{eq:flow}) with the CTVF through the forward Euler method. For $k=0,...,K-1$, the surface points are updated by $x_{k+1}=x_k+hv_k(x_k;a)$, where $K$ is the total integration steps and $h=T/K$ is the step size with $T=1$. For each step $k$, we only need to recompute the attention maps $\mathbf{p}(hk,a)$ and update the CTVF $v_k(x_k;a)$ accordingly by Eq.~(\ref{eq:ctvf}). CoTAN only integrates a single CTVF which saves considerable runtime compared to integrating multiple SVFs directly as~\cite{gupta2020nmf,lebrat2021corticalflow,santa2022cfpp}.
\newline

\begin{figure}[t]
\centering
\includegraphics[width=1.0\textwidth]{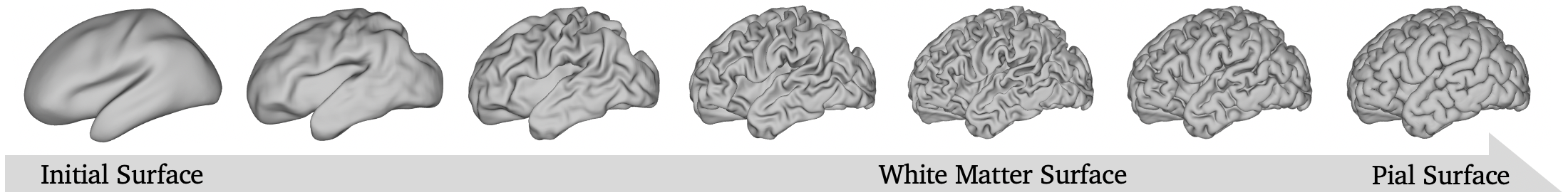}
\caption{Diffeomorphic deformation from an initial template mesh to cortical surfaces.} \label{fig:deform}
\end{figure}

\noindent\textbf{Neonatal Cortical Surface Reconstruction. }
We train two CoTAN models on the dHCP neonatal dataset~\cite{edwards2022dhcp} to 1) deform an initial surface into a white matter surface and to 2) expand the white matter surface into a pial surface as shown in Figure~\ref{fig:deform}. We use the same initial surface (leftmost in Figure~\ref{fig:deform}) for all subjects, which is created by iteratively applying Laplacian smoothing to a Conte-69 surface atlas~\cite{glasser2011conte}. We generate pseudo ground truth (GT) surfaces by the dHCP structural neonatal pipeline~\cite{makropoulos2018dhcp}, which has been fully validated through quality control performed by clinical experts.

For white matter surface reconstruction, we consider loss functions that have been widely used in previous work~\cite{bongratz2022vox2cortex,wang2018pixel2mesh,wickramasinghe2020voxel2mesh}: the Chamfer distance loss $\mathcal{L}_{cd}$ computes the distance between two point clouds, the mesh Laplacian loss $\mathcal{L}_{lap}$ regularizes the smoothness of the mesh, and the normal consistency loss $\mathcal{L}_{nc}$ constrains the cosine similarity between the normals of two adjacent faces. The final loss is weighted by $\mathcal{L}=\mathcal{L}_{cd}+\lambda_{lap}\mathcal{L}_{lap}+\lambda_{nc}\mathcal{L}_{nc}$.
We train CoTAN in two steps for white matter surface extraction. First, we pre-train the model using relatively large weights $\lambda_{lap}$ and $\lambda_{nc}$ for regularization. The Chamfer distance is computed between the vertices of the predicted and pseudo-GT surfaces. These ensure that the initial surface can be deformed robustly during training. Then, we fine-tune CoTAN using small weights to increase geometric accuracy. The distances are computed between 150k uniformly sampled points on the surfaces. 

For pial surface reconstruction, we follow~\cite{ma2022cortexode,ma2021pialnn} and use the pseudo-GT white matter surfaces as the input for training. Then the MSE loss $\mathcal{L}=\sum_i\|\hat{x}_i-x^*_i\|^2$ can be computed between the vertices of predicted and pseudo-GT pial meshes. No point matching is required since the pseudo-GT white matter and pial surfaces have the same mesh connectivity. Therefore, the MSE loss provides stronger supervision than the Chamfer distance, while the latter is prone to mismatching the points in narrow sulci, resulting in mesh self-intersections.

\section{Experiments}
\noindent\textbf{Implementation Details.} We evaluate CoTAN on the third release of dHCP neonatal dataset~\cite{edwards2022dhcp} (\url{https://biomedia.github.io/dHCP-release-notes/}), which includes 877 T2-weighted (T2w) brain MRI scanned from newborn infants at PMA between 27 to 45 weeks. The MRI images are affinely aligned to the MNI152 space and clipped to the size of $112\times224\times160$ for each brain hemisphere. The dataset is split into 60/10/30\% for training/validation/testing. 

For the CoTAN model, we set the resolution $R$=3 and the number of SVFs $M$=4 for each resolution. For integration, we set the total number of steps to $K$=50 with step size $h$=0.02. We re-mesh the initial mesh to 140k vertices, of which the coordinates are normalized to $[-1,1]$. We use the Adam optimizer for training. For the white matter surface, we first pre-train CoTAN for 100 epochs using a learning rate of $\gamma$=$10^{-4}$ and weights $\lambda_{lap}$=$0.5$, $\lambda_{nc}$=5$\times$$10^{-4}$ for the loss function. Then we fine-tune for 100 epochs using smaller weights $\lambda_{lap}$=$0.1$ and $\lambda_{nc}$=$10^{-4}$ with $\gamma$=2$\times$$10^{-5}$. For the pial surface, we set the maximum channel size of CoTAN as 32 to avoid overfitting and train for 200 epochs with $\gamma$=$10^{-4}$. We only consider left brain hemisphere in the experiments. All experiments are conducted on a Nvidia RTX3080 GPU with 12GB memory.
\newline

\begin{table}[t] \scriptsize
\centering
\caption{Comparative results of neonatal cortical surface reconstruction on the dHCP dataset. The geometric accuracy (ASSD and HD90) and mesh quality (the ratio of SIFs) are reported for white matter and pial surfaces. Smaller values mean better results.\newline 
*CoTAN (ours) shows significant improvement ($p$<0.05) compared to baselines.}
\label{tab:comparison}
\centerline{
\begin{tabular}{l|ccc|ccc}
\toprule
& \multicolumn{3}{c|}{White Matter Surface} & \multicolumn{3}{c}{Pial Surface} \\
Method & 
\multicolumn{1}{c}{ASSD (mm)} & \multicolumn{1}{c}{HD90 (mm)} & \multicolumn{1}{c|}{SIF ($\%$)} &
\multicolumn{1}{c}{ASSD (mm)} & \multicolumn{1}{c}{HD90 (mm)} & \multicolumn{1}{c}{SIF ($\%$)} \\
\midrule
CoTAN  &
\textbf{0.107}$\pm$\textbf{0.026} & \textbf{0.217}$\pm$\textbf{0.076} & \textbf{0.001}$\pm$\textbf{0.004} & 
\textbf{0.121}$\pm$\textbf{0.029} & \textbf{0.259}$\pm$\textbf{0.075} & \textbf{0.071}$\pm$\textbf{0.034}\\
CortexODE & 
0.109$\pm$0.052~ & 0.231$\pm$0.326~ & \textbf{0.001}$\pm$\textbf{0.002} & 
0.134$\pm$0.052* & 0.306$\pm$0.358* & 0.221$\pm$0.114*\\
CFPP & 
0.118$\pm$0.028* & 0.241$\pm$0.085* & 0.075$\pm$0.057* & 
0.124$\pm$0.031* & 0.273$\pm$0.086* & 2.457$\pm$1.003*\\
CorticalFlow & 
0.122$\pm$0.029* & 0.247$\pm$0.080* & 0.048$\pm$0.032* & 
0.157$\pm$0.031* & 0.331$\pm$0.089* & 9.798$\pm$2.902*\\
Vox2Cortex & 
0.115$\pm$0.035* & 0.233$\pm$0.110* & 0.253$\pm$0.169* & 
0.130$\pm$0.039* & 0.291$\pm$0.141* & 14.366$\pm$2.262*\\
DeepCSR & 
0.129$\pm$0.047* & 0.276$\pm$0.211* & --- & 
0.299$\pm$0.070* & 1.214$\pm$0.337* & ---\\
\bottomrule
\end{tabular}}
\end{table}

\begin{table}[t] 
\scriptsize 
\centering
\begin{minipage}{.53\linewidth}
\centering
\caption{Comparative results of runtime, GPU memory cost, and the number of model parameters for both training and testing.}
\label{tab:efficiency}
\begin{tabular}{l|cc|cc|c}
\toprule
&  \multicolumn{2}{c|}{Runtime} & \multicolumn{2}{c|}{GPU (GB)} & Model\\
Method & \multicolumn{1}{c}{Train} & \multicolumn{1}{c|}{Test} & \multicolumn{1}{c}{Train} & \multicolumn{1}{c|}{Test}& 
\multicolumn{1}{c}{\#Param} \\
\midrule
CoTAN (Ours) &
57.9h & \textbf{0.21}s & 8.71 & 4.05 & 2.47M \\
CortexODE  
& 51.6h & 1.88s & 9.24 & 4.61 & 1.99M \\
CFPP  
& 131.5h & 0.57s & 9.86 & 3.56 & 1.03M \\
CorticalFlow  
& 105.7h & 0.51s & 9.12 & 3.56 & 1.03M \\
Vox2Cortex  
& 63.7h & 1.48s & 6.96 & 6.26 & 6.54M  \\
DeepCSR  
& \textbf{15.1}h & 10.69s & \textbf{5.26} & \textbf{2.33} & 4.65M \\
\bottomrule
\end{tabular}
\end{minipage}
\hfill
\begin{minipage}{.45\linewidth}
\centering
\caption{The results of ablation experiments for CoTAN on white matter surface reconstruction.}
\label{tab:ablation}
\begin{tabular}{l|cc}
\toprule
Method & 
ASSD (mm) & HD90 (mm) \\
\midrule
CoTAN (Ours)&
\textbf{0.107}$\pm$\textbf{0.026} & \textbf{0.217}$\pm$\textbf{0.076}\\
Pre-train & 0.116$\pm$0.029 & 0.238$\pm$0.095 \\
$R$=1 & 0.112$\pm$0.027 & 0.232$\pm$0.081 \\
$M$=1 & 0.111$\pm$0.030 & 0.231$\pm$0.097 \\
SVF ($a$=$t$=0) & 0.138$\pm$0.037 & 0.291$\pm$0.109 \\
CVF ($t$=0) & 0.135$\pm$0.036 & 0.285$\pm$0.111 \\
TVF ($a$=0) & 0.108$\pm$0.028 & 0.222$\pm$0.090 \\
\bottomrule
\end{tabular}
\end{minipage}
\end{table}

\begin{figure}[t]
\centering
\includegraphics[width=1.0\textwidth]{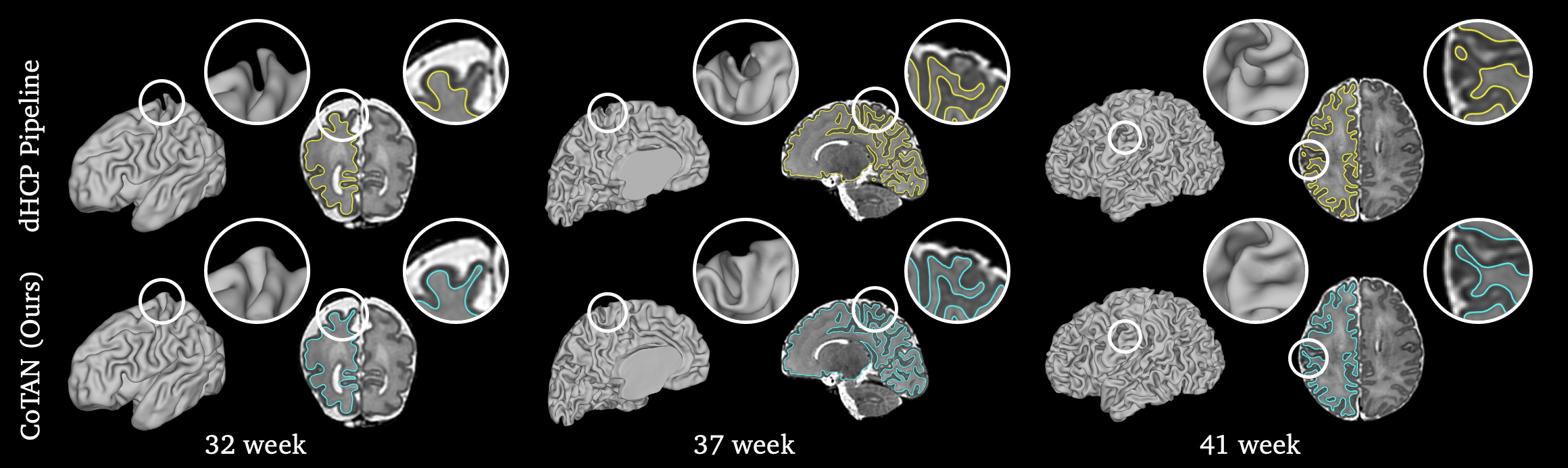}
\caption{Visualization of neonatal cortical surfaces generated by the dHCP structural pipeline~\cite{makropoulos2018dhcp} and CoTAN for different ages. CoTAN shows better anatomical accuracy as highlighted in both surface meshes and corresponding brain MRI images.} \label{fig:compare}
\end{figure}

\noindent\textbf{Comparative Results.} We compare the performance of CoTAN with existing learning-based cortical surface extraction approaches including CortexODE~\cite{ma2022cortexode}, CorticalFlow++ (CFPP)~\cite{santa2022cfpp}, CorticalFlow~\cite{lebrat2021corticalflow}, Vox2Cortex~\cite{bongratz2022vox2cortex} and DeepCSR~\cite{cruz2021deepcsr}. We employ the fast topology correction~\cite{ma2022cortexode} for DeepCSR.  CoTAN can guarantee spherical topology and the Euler number is 2 for all predicted surfaces. 

\noindent---\textit{Geometric accuracy}: We measure the geometric accuracy by commonly used metrics~\cite{bongratz2022vox2cortex,cruz2021deepcsr,ma2022cortexode}: average symmetric surface distance (ASSD) and 90th percentile of Hausdorff distance (HD90). The distances are computed between uniformly sampled 100k points on the predicted and pseudo-GT surfaces. For fair comparison, the predicted cortical meshes have around 140k vertices for all baseline approaches. Note that CoTAN and \cite{cruz2021deepcsr,lebrat2021corticalflow,santa2022cfpp} can generalize on high-resolution meshes with up to 600k vertices (see Appendix). We conduct a paired t-test to examine the statistical significance. As reported in Table~\ref{tab:comparison}, CoTAN achieves significantly superior geometric accuracy ($p$-value<0.05) compared to all state-of-the-art baselines, except for the white matter surfaces of CortexODE.

\noindent---\textit{Mesh quality}: We evaluate the mesh quality by the ratio of self-intersecting faces (SIF) as shown in Table \ref{tab:comparison}. Note that due to the narrower sulcal gaps of neonatal cortex compared to adult~\cite{jack2008adni,marcus2007oasis} (see Figure \ref{fig:sulci}), all baseline methods produce more SIFs than reported in their original papers. Except that DeepCSR produces no SIFs since it uses the Marching Cubes algorithm, CoTAN achieves the best mesh quality with only 0.001\% SIFs in white matter surfaces and 0.071\% SIFs in pial surfaces. These remaining SIFs are likely introduced by the discretization of triangular mesh representation and ODE integration. 

CoTAN produces fewer SIFs for three reasons. Firstly, CoTAN employs diffeomorphic deformation, while the non-diffeomorphic Vox2Cortex creates 14\% SIFs. We further set the integration steps $K$=5 for CoTAN so that the deformation is no longer diffeomorphic. The SIFs of pial surfaces are increased to 2.99$\pm$1.19\%. Secondly, CoTAN reconstructs the pial surface by expanding the input white matter surface. It is difficult to avoid collisions during the
deformation from a smooth template into deep sulci, \emph{e.g.}, in Vox2Cortex and CorticalFlow. Lastly, CoTAN uses MSE loss rather than Chamfer distance for pial surface extraction, which alleviates the mismatch between points in the narrow sulci.

\noindent---\textit{Computational efficiency}: We report the runtime and GPU memory cost for both training and testing, as well as the number of learnable parameters for CoTAN and all baseline approaches in Table \ref{tab:efficiency}. The runtime includes both model inference and post-processing. CoTAN only requires 0.21 seconds to extract cortical surfaces for each hemisphere, which is 2$\times$ faster than the best baseline.
CoTAN can be trained end-to-end and reduces training time by 46\% compared to CorticalFlow and CFPP, which have to train three U-Nets consecutively to parameterize three SVFs for a single surface. Although CoTAN uses relatively more parameters to learn surface deformations, it is still memory efficient which only costs 8.7GB and 4GB GPU memory for training and testing.
\newline

\noindent\textbf{Ablation Study.} We conduct ablation studies on CoTAN and evaluate the geometric accuracy on the white matter surface reconstruction. First, without fine-tuning, the geometric errors increase by 8\% as reported in Table~\ref{tab:ablation}. Then we consider CoTAN with single resolution ($R$=1) or only predict a single SVF ($M$=1) for each resolution. The geometric distances increase in both cases.

Next, we examine the effectiveness of the CTVF $v_t(x;a)$ by fixing the input time $t$=0 or age $a$=0. We train CoTAN models to predict the TVF ($a$=0), CVF ($t$=0) and SVF ($a$=$t$=0), which are degraded from the CTVF. Table~\ref{tab:ablation} shows that the SVF increases geometric errors by 30\% due to its limited representation capacity. The CVF increases accuracy slightly by learning conditional deformations adaptive to the age of neonates. The TVF exploits temporal information to model a wider range of deformations and effectively improves the performance.

Lastly, we train a U-Net to predict $R$$\times$$M$ SVFs and integrate them directly without attention. Since the gradients are backpropagated through all SVFs, it requires >140 hours training time which is 2.4$\times$ slower than our attention-based CoTAN. The model is also sensitive to small updates, which can affect all SVFs. This results in exploding gradients which we have observed in the training, whereas CoTAN can be trained robustly by integrating a single CTVF.
\newline

\begin{figure}[t]
\centering
\begin{minipage}{.54\linewidth}
\centering
\includegraphics[width=1.0\textwidth]{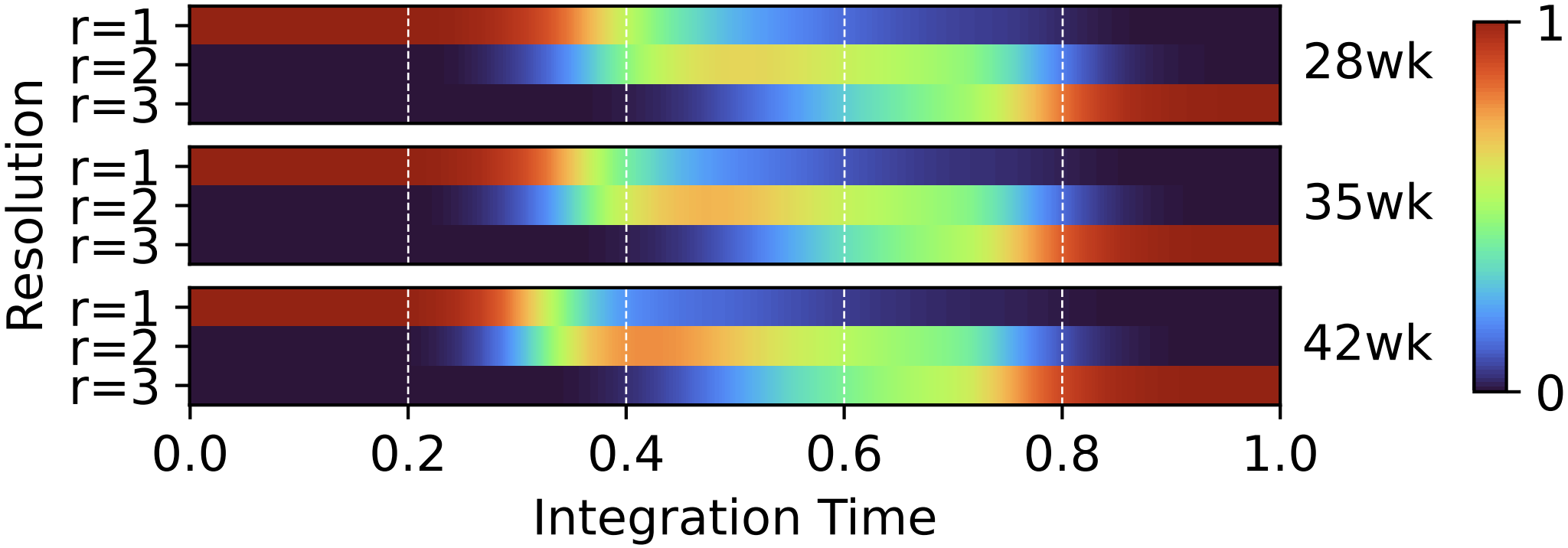}
\end{minipage}
\hfill
\begin{minipage}{.44\linewidth}
\includegraphics[width=1.0\textwidth]{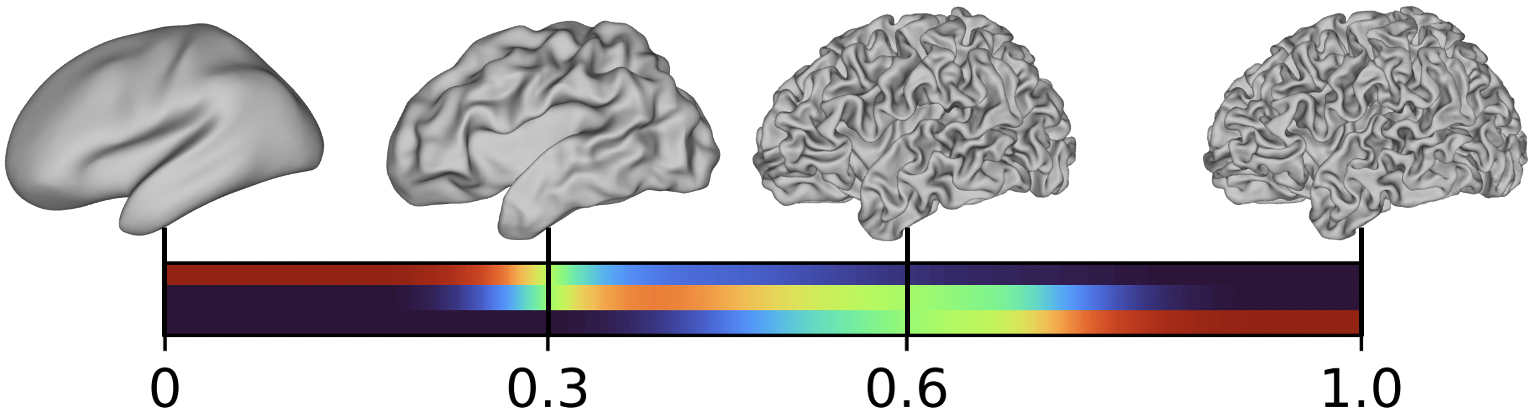}
\end{minipage}
\caption{Visualization of attention maps predicted by CoTAN for white matter surface reconstruction. Left: conditional temporal attention maps $\mathbf{p}^r(t,a)$ at different time and ages. Right: coarse-to-fine white matter surface deformations learned by CoTAN.} \label{fig:attention}
\end{figure}

\noindent\textbf{Attention Maps. } 
We explore the attention maps $\mathbf{p}^{r,m}(t,a)$ learned by CoTAN. We define $\mathbf{p}^r=\sum_{m=1}^M\mathbf{p}^{r,m}$ to reflect the importance of the SVFs at each resolution level $r$=$1,...,R$, where $R=3$ and larger $r$ means higher resolution. Figure \ref{fig:attention} visualizes the attention maps $\mathbf{p}^r(t,a)$ for white matter surface reconstruction for integration time $t\in[0,1]$ and age $a\in\{28,35,42\}$. It shows that the attention maps focus on low resolution ($r$=1) at the beginning of the integration, and then attends to high resolution ($r$=3) when $t$$\rightarrow$1. Furthermore, Figure \ref{fig:attention} (Right) shows that an initial white matter surface deforms into a coarse shape for $t$$\leq$0.3 and learns fine details for $t$$\geq$0.6. This matches the attention maps and demonstrates that CoTAN learns coarse-to-fine deformations automatically without any supervision on the intermediate deformations. Additionally, Figure \ref{fig:attention} shows that CoTAN pays more attention to low resolution for younger subjects (28 week) whose brains have not fully developed yet. More deformations at higher resolutions ($r$$\geq$2) are required for older neonates ($\geq$35 week) with highly folded cortex. 
\newline

\noindent\textbf{Discussion.} One limitation of our experiments is that we only train and evaluate CoTAN based on the pseudo-GT generated by the dHCP structural pipeline~\cite{makropoulos2018dhcp}. Previous approaches~\cite{bongratz2022vox2cortex,cruz2021deepcsr,ma2022cortexode} have been validated by the test-retest experiments. However, this is infeasible for neonates, whose brain develops rapidly even within a short period. To verify the superior anatomical accuracy of CoTAN, we provide qualitative comparison between the pseudo-GT and CoTAN as visualized in Figure \ref{fig:compare}. It shows that CoTAN can effectively mitigate corruptions introduced by the dHCP pipeline for neonatal subjects at different ages. In addition, the dHCP pipeline requires 4 hours to process a single subject~\cite{makropoulos2018dhcp}, while CoTAN extracts cortical surfaces in only 0.21 seconds for each brain hemisphere.

\section{Conclusion}
In this work, we propose CoTAN for diffeomorphic neonatal cortical surface reconstruction. CoTAN employs an attention mechanism to combine multiple SVFs to a CTVF, which outperforms existing baselines in geometric accuracy and mesh quality. CoTAN can also be extended and applied to extract adult cortical surfaces conditioned on the age, gender or pathological information of the subjects. Our future work will integrate CoTAN into a learning-based pipeline for universal cortical surface analysis across all age groups.\newline

\noindent\textbf{Acknowledgements. }
This work was supported by the President’s PhD Scholarship at Imperial College London. Support was also received from the ERC project MIA-NORMAL 101083647 and ERC project Deep4MI 884622. Data were provided by the developing Human Connectome Project, KCL-Imperial-Oxford Consortium funded by the ERC under the European Union Seventh Framework Programme (FP/2007-2013) / ERC Grant Agreement no. [319456]. We are grateful to the families who generously supported this trial.

%
%
\bibliographystyle{splncs04}
\bibliography{paper1290}

\end{document}